\documentclass[aps,pre,twocolumn,showpacs]{revtex4}
\usepackage{amssymb,amsmath}
\usepackage[dvips]{graphicx}

\begin{document}

\title{Onset of Collective Oscillation in Chemical Turbulence
under Global Feedback}

\author{Yoji Kawamura}
\email{kawamura@ton.scphys.kyoto-u.ac.jp}

\author{Yoshiki Kuramoto}

\affiliation{Department of Physics, Graduate School of Sciences,
Kyoto University, Kyoto 606-8502, Japan}

\date{May 30, 2003}


\begin{abstract}
Preceding the complete suppression of chemical turbulence by means
of global feedback, a different universal type of transition, which
is characterized by the emergence of small-amplitude collective
oscillation with strong turbulent background, is shown to occur at
much weaker feedback intensity.
We illustrate this fact numerically in combination with a phenomenological
argument based on the complex Ginzburg-Landau equation with global feedback.
\end{abstract}

\maketitle

\section{Introduction} \label{sec:intro}

Chemical turbulence in oscillatory reaction-diffusion systems
can be completely suppressed by means of global feedback
~\cite{ref:battogtokh96,ref:battogtokh97,ref:kim01}.
Theoretically, this fact was found in the complex Ginzburg-Landau equation
~\cite{ref:battogtokh96,ref:battogtokh97}, i.e., the normal form
of oscillatory reaction-diffusion systems near the supercritical
Hopf bifurcation point~\cite{ref:kuramoto84}.
Recent experiments on catalytic CO oxidation on Pt surface demonstrated
the same fact, revealing also a variety of wave patterns caused by the
effects of global delayed feedback~\cite{ref:kim01,ref:bertram03-1}.
A theoretical model for this reaction system reproduced similar
behavior~\cite{ref:kim01,ref:bertram03-2}.

In the present paper, we show that yet another transition of
universal nature can occur at a certain feedback intensity which
is much weaker than the critical intensity associated with the
complete suppression of turbulence.
The new type of transition is characterized by the emergence of small-amplitude
collective oscillation out of the strongly turbulent medium without
long-range phase coherence.
When the collective oscillation appeared, the system remains strongly turbulent,
while the effective damping rate of the uniform mode (i.e., the mean field) shows
a change of sign from positive to negative.
Thus, the transition is interpreted as a consequence of a complete cancellation
of the effective damping of the mean field with the effect of its growth
produced by the global feedback.

In Sec.~\ref{sec:langevin}, we start with the complex
Ginzburg-Landau equation (CGL) with global feedback.
Then we derive phenomenologically a nonlinear Langevin equation governing
the mean field in the form of a noisy Stuart-Landau equation (SL).
In order to clarify the nature of the transition of our concern,
some numerical results for the one- and two- dimensional CGL will
be compared in Secs.~\ref{sec:results} and \ref{sec:two},
respectively, with analytical results obtained from the noisy SL. 
Concluding remarks will be given in the final section.

\section{Langevin equation for turbulent CGL as an effective equation}
\label{sec:langevin}

One-dimensional complex Ginzburg-Landau equation with global feedback
is given by~\cite{ref:battogtokh96,ref:battogtokh97}
\begin{equation}
\partial_t W = W+\left(1+i c_1\right)\partial^2_x W
-\left(1+i c_2\right)\left|W\right|^2 W+\mu\bar{W},
\label{eq:cgl}
\end{equation}
\begin{equation}
\bar{W}\left(t\right) = \frac{1}{L}\int_0^{L}W\left(t, x\right)dx,
\label{eq:gf}
\end{equation}
where $W$ is a complex field, and $L$ is the system size which
is supposed to be sufficiently large.
The intensity $\mu$ of the  global feedback is generally a complex number.
It is known, however, that a suitable tuning of the delay time in the
feedback in the original system can control the phase of this parameter
~\cite{ref:battogtokh96,ref:battogtokh97}.
For the sake of simplicity, therefore, we shall confine our present analysis
to the case of real $\mu$, which corresponds to the situation where the delay
time in the feedback is fixed at a certain value but the feedback intensity
is allowed to vary.
A brief comment will be made on the case of complex $\mu$ in the final section.

We first consider the system without feedback ($\mu=0$), i.e., the usual
one-dimensional CGL~\cite{ref:aranson02,ref:shraiman92}.
As is well known, uniform oscillations are linearly unstable and
turbulence develops when the Benjamin-Feir instability condition
\begin{equation}
1+c_1 c_2<0
\label{eq:bf}
\end{equation}
is satisfied.
In what follows, we will fix the parameters $c_1$ and $c_2$ as $c_1=2.0$
and $c_2=-2.0$ so that the system may stay well within the turbulent regime.
We confirmed that under this condition no collective oscillation exists,
i.e., $\bar{W}$ is randomly fluctuating on a ``microscopic'' scale around
the zero value without perceptible systematic motion.
The core of our argument developed below depends little on the choice of
parameter values as far as the condition~(\ref{eq:bf}) is well satisfied. 

\begin{figure}
\centering
\includegraphics[width=8cm]{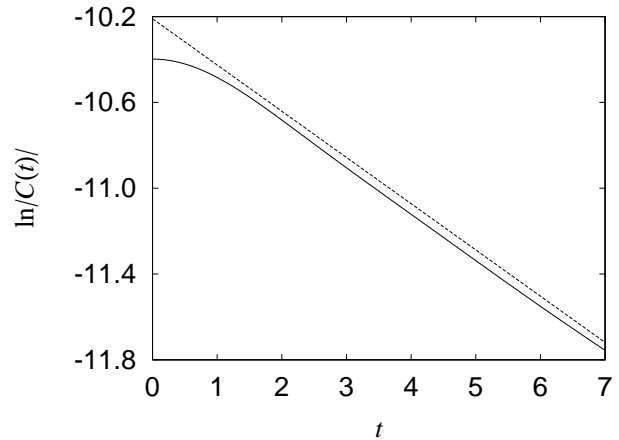}
\caption{Numerically observed exponential decay of the time-correlation
function of the mean field in semi-logarithmic scales. The effective
damping coefficient $\gamma$, i.e., the mean tangent (the broken line)
of the curve with the initial transient excluded, is estimated to be $0.22$.
The system size is $N=200000$.}
\label{fig:decay}
\end{figure}

It is known that if the turbulence is sufficiently strong, which
is actually the case under the above parameter condition, the
system exhibits extensive chaos characterized by the property
~\cite{ref:egolf94}
\begin{equation}
D_{\rm f} \propto L,
\label{eq:ec}
\end{equation}
where $D_{\rm f}$ is the Lyapunov dimension of the high-dimensional
chaotic attractor describing the turbulence.
Extensive chaos implies that the system can be imagined as composed
of a large number of cells of equal size such that the fluctuations
of some variables associated with the individual cells about their
mean value are  statistically independent from cell to cell.
Thus, the fluctuations of a macro-variable, i.e., a variable given by
a simple sum of cell variables over the entire system, are expected
to obey the central limit theorem.
In particular, in the absence of long-range order, the characteristic
amplitude of $\bar{W}$ will scale like $1/\sqrt{L}$ for large system size.
If we represent our continuous oscillatory medium with a long array of $N$
oscillators with sufficiently small but fixed separation between neighboring
oscillators, which we actually do in numerical simulations to be described below,
we expect asymptotically the property
\begin{equation}
\bar{W}=\frac{1}{N}\sum_{j=1}^N W_j=O\left(\frac{1}{\sqrt{N}}\right),
\label{eq:clt}
\end{equation}
where $W_j$ is the complex amplitude of the $j$-th oscillator in the array.
The simple nature of the extensive variable $\bar{W}$ mentioned above
also implies that its time-correlation function defined by
\begin{equation}
C\left(t\right) = \left\langle\bar{W}^{\ast}(0)\bar{W}(t)\right\rangle
\equiv\lim_{T\to\infty}\frac{1}{T}\int_0^T
\bar{W}^{\ast}(s)\bar{W}(s+t)ds
\label{eq:cor}
\end{equation}
obeys a simple exponential decay law for large $t$, or
\begin{equation}
C(t) \propto \exp\left(-\gamma t\right).
\label{eq:exp}
\end{equation}
The effective damping coefficient $\gamma$ is a quantity which
could only be determined from the statistical mechanics of turbulent
fluctuations which is still far from being established.
The above decay law is confirmed from our numerical simulation.
Figure~\ref{fig:decay} shows numerically calculated $C(t)$ from
which the exponential law with $\gamma\simeq 0.22$ is confirmed except
for some initial transient.
In this numerical simulation, and in all numerical simulations which
follow, we applied an explicit Euler integration scheme with a constant
time step $\varDelta t\leq 0.01$ and a fixed grid size $\varDelta x = 0.5$,
adopting periodic boundary conditions.
The system size $N$ used ranges from $400$ to $200000$.

We now introduce global feedback and study its effects on the dynamics
of the mean field. Let the complex amplitude be decomposed into Fourier
series as
\begin{equation}
W = \sum_{k=-\infty}^{\infty} \tilde{W_k}\, e^{i q_k x},
\label{eq:ft}
\end{equation}
where $q_k = 2\pi k/L$ ($k$ is an integer), and the Fourier amplitudes
are defined by
\begin{equation}
\tilde{W}_k = \frac{1}{L}\int_0^L W\, e^{-i q_k x}dx.
\label{eq:fa}
\end{equation}
The uniform amplitude $\tilde{W}_0$, which is identical with the mean
field $\bar{W}$ by definition, obeys the equation
\begin{equation}
\dot{\tilde{W_0}} = \tilde{W}_0
-\left(1+i c_2\right)\sum_{k_1, k_2}
\tilde{W}_{k_1}\tilde{W}_{k_2}\tilde{W}_{k_1+k_2}^{\ast}
+\mu\tilde{W}_0.
\label{eq:um}
\end{equation}
One may wish to obtain an equation for the mean field in a closed form,
which would be a stochastic equation of the nonlinear Langevin type.
However, deriving such an equation would be a formidable statistical
mechanical problem, so that in the present paper we will content
ourselves by simply assuming phenomenologically that the effective
exponential decay of the mean field as is seen in Fig.~\ref{fig:decay}
is a result of renormalization of the linear coefficient in Eq.~(\ref{eq:um})
by the nonlinear mode-coupling term.
If we can neglect non-Markovian effects, the result of such renormalization
will generally take the form of a nonlinear Langevin equation
\begin{equation}
\dot{\bar{W}} =
\left[\left(\mu-\gamma\right)+i\omega\right]\bar{W}
+{\cal N}\left(\bar{W}\right)+f(t),
\label{eq:le}
\end{equation}
where a nonlinear term ${\cal N}(\bar{W})$ with unknown specific form
has been included, and $f(t)$ represents random force with vanishing mean.

The analysis of Eq.~(\ref{eq:le}) developed in the following section is
based on the simplifying assumption that $f(t)$ is white Gaussian with
the only non-vanishing second moment given by
\begin{equation}
\left\langle f(t)f^{\ast}(t')\right\rangle=4\Gamma\delta(t-t').
\label{eq:n2}
\end{equation}
The Gaussian nature of the random force seems to hold due to the
aforementioned extensive nature of $\bar{W}$.
The white-noise assumption also seems valid because we are particularly
concerned with the situation near the transition point where the
characteristic time scale of $\bar{W}$ becomes very long.
We should also note that the damping coefficient $\gamma$ has been assumed
to be unchanged when the global feedback is introduced, which is actually the
property confirmed by our numerical experiments at least for real $\mu$.
Our final remark is that in the above argument about Eq.~(\ref{eq:le})
we did not explicitly refer to spatial dimension.
Therefore, there seems to be no reason why Eq.~(\ref{eq:le}) should
not be applied to systems of two or higher dimensions.
 
\begin{figure}
\centering
\includegraphics[width=8cm]{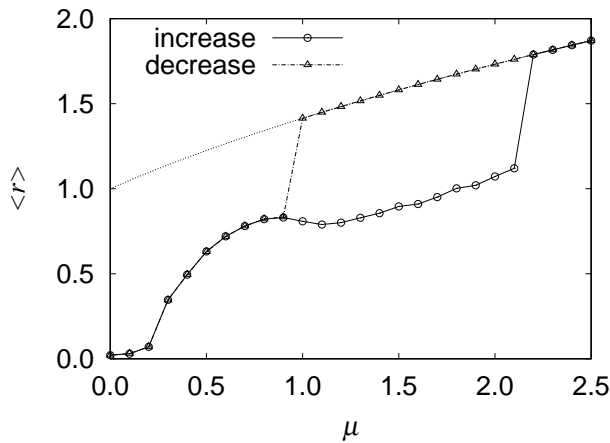}
\caption{Long-time averages of the mean field amplitude
$\langle r\rangle$ as a function of the feedback intensity $\mu$
over a wide range of $\mu$. The dotted line corresponds to
uniform oscillations. The system size is $N=10000$.}
\label{fig:over}
\end{figure}

Equation~(\ref{eq:le}) tells that a transition occurs when
the global feedback intensity $\mu$ becomes equal to the effective
damping coefficient $\gamma$ of the mean filed.
We denote this value of $\mu$ as $\mu_{\rm c}$, or
\begin{equation}
\mu_{\rm c} = \gamma.
\label{eq:muc}
\end{equation} 
Figure~\ref{fig:over} shows numerically obtained long-time averages
of the mean field amplitude, denoted as $\langle r\rangle$, as a function
of the feedback intensity $\mu$ over a wide range of $\mu$.
Although not very clear from these data, there is an indication of transition from
vanishing to non-vanishing $\langle r\rangle$ at some value of $\mu$ much smaller
than those giving rise to a hysteresis between turbulent and non-turbulent
uniform states.
In fact, the whole analysis in the rest of this paper is devoted to finding
out unambiguous evidence for the existence of a transition through a closer
investigation of small-$\mu$ region.

\begin{figure}
\centering
\begin{tabular}{c c c}
 & (a) & (b) \\[1mm]
\rotatebox{90}{\hspace{1.2cm}{\Large time $\rightarrow$}} &
\includegraphics[width=4cm]{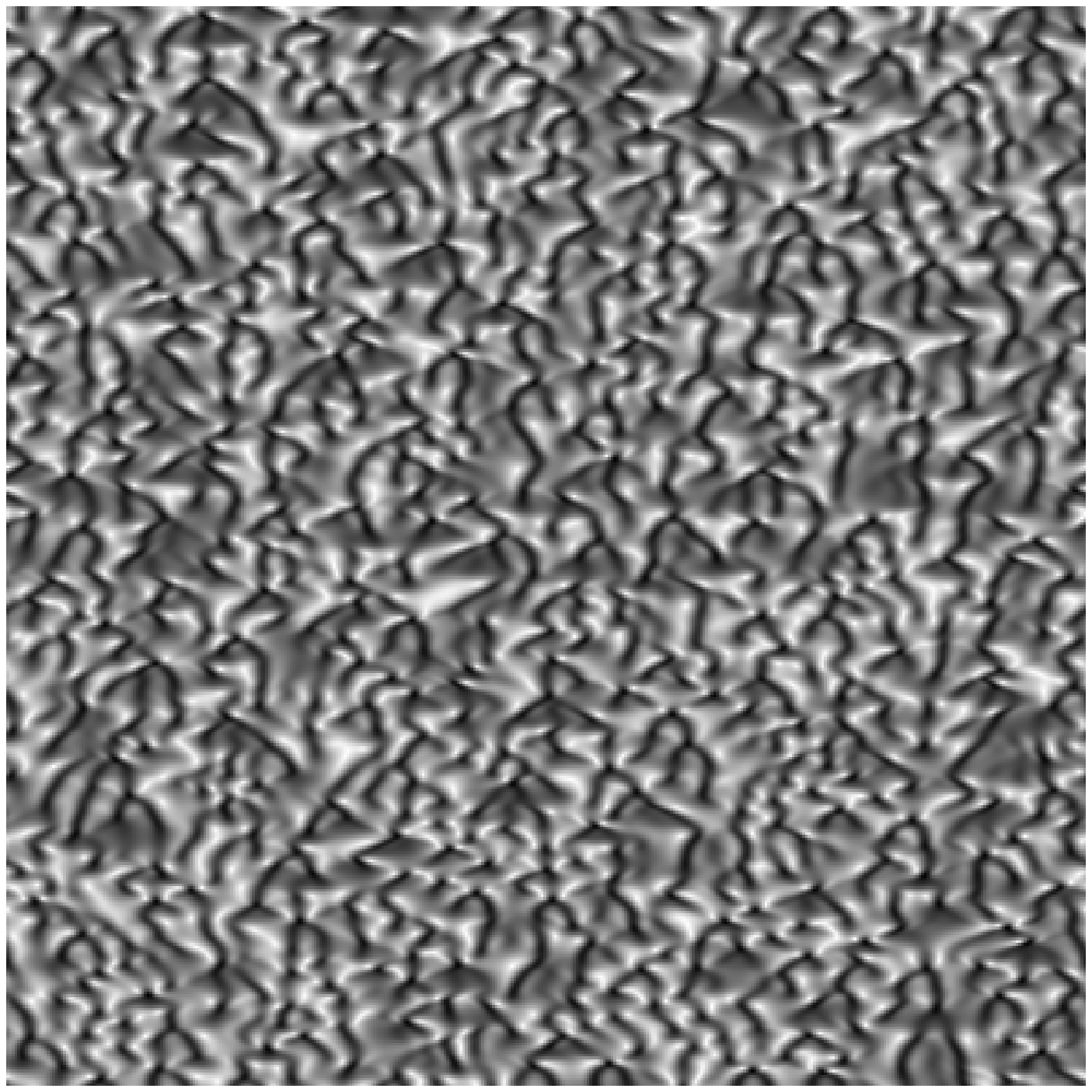} &
\includegraphics[width=4cm]{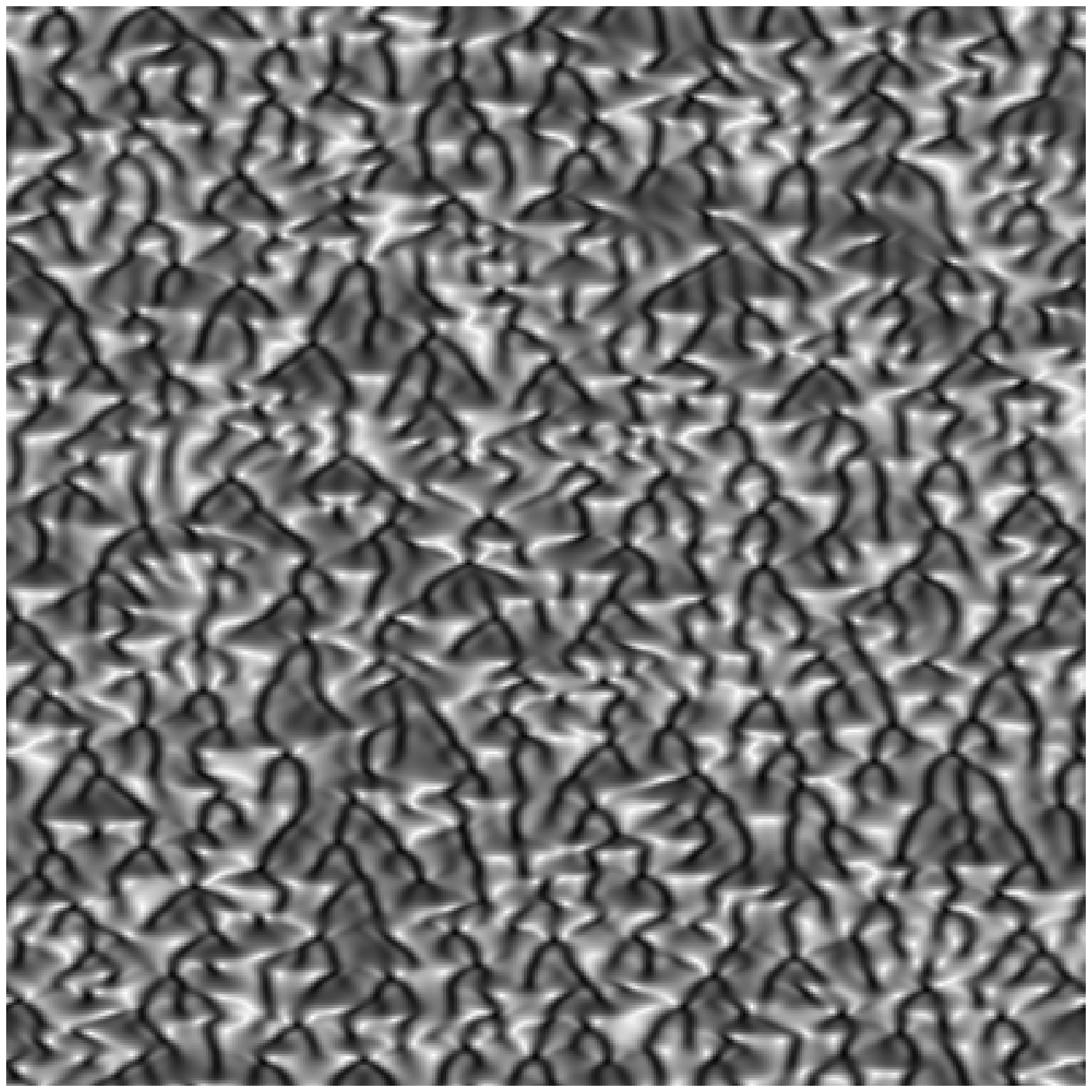}
\end{tabular}
\caption{Space (horizontal) - time (vertical) plot of the complex
amplitude modulus $|W|$ for $\mu=0.1$ (a) and $\mu=0.3$ (b).
The system size is $N=400$.}
\label{fig:stp}
\end{figure}

\begin{figure}
\centering
\includegraphics[width=8cm]{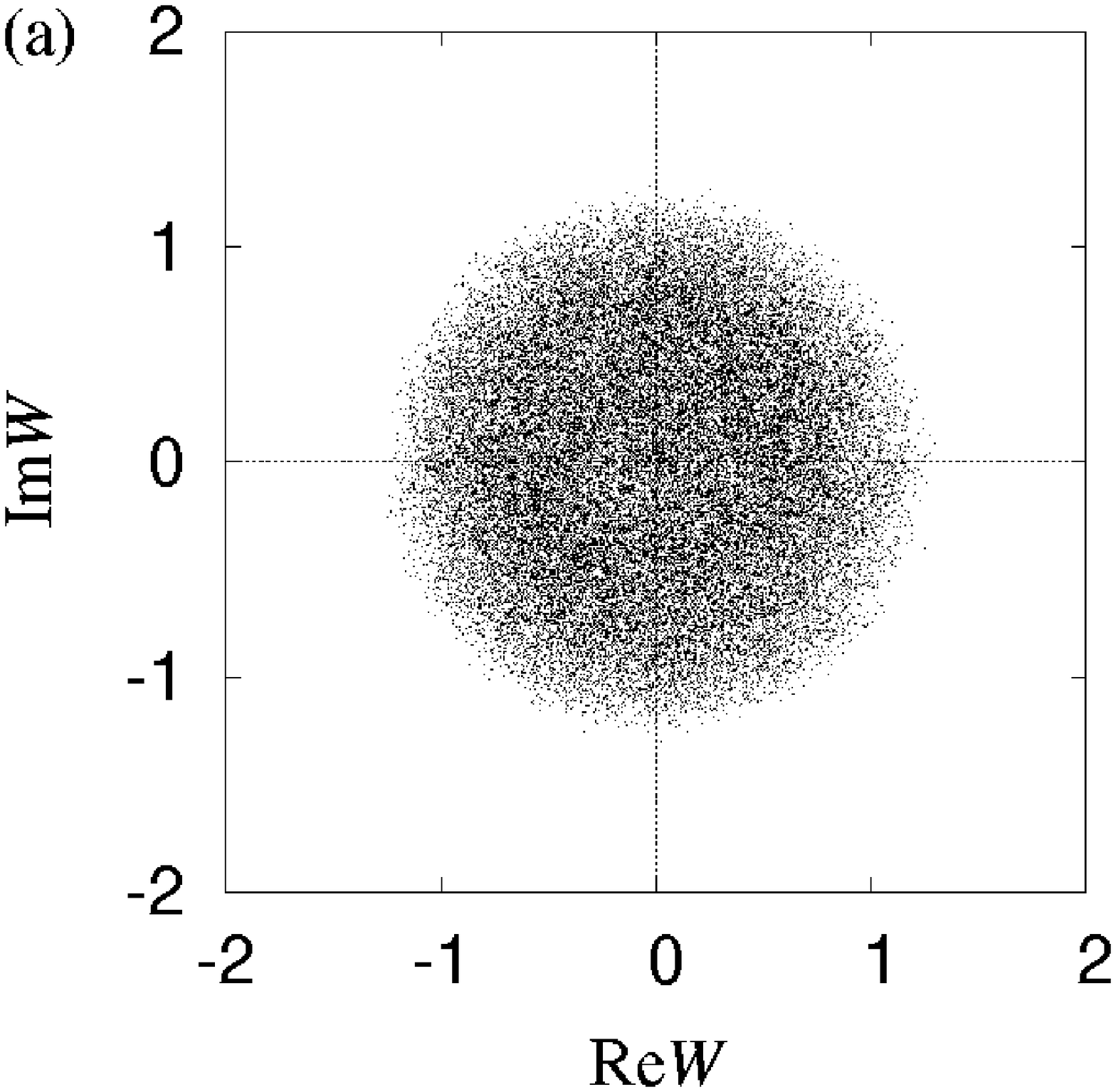} \\
\includegraphics[width=8cm]{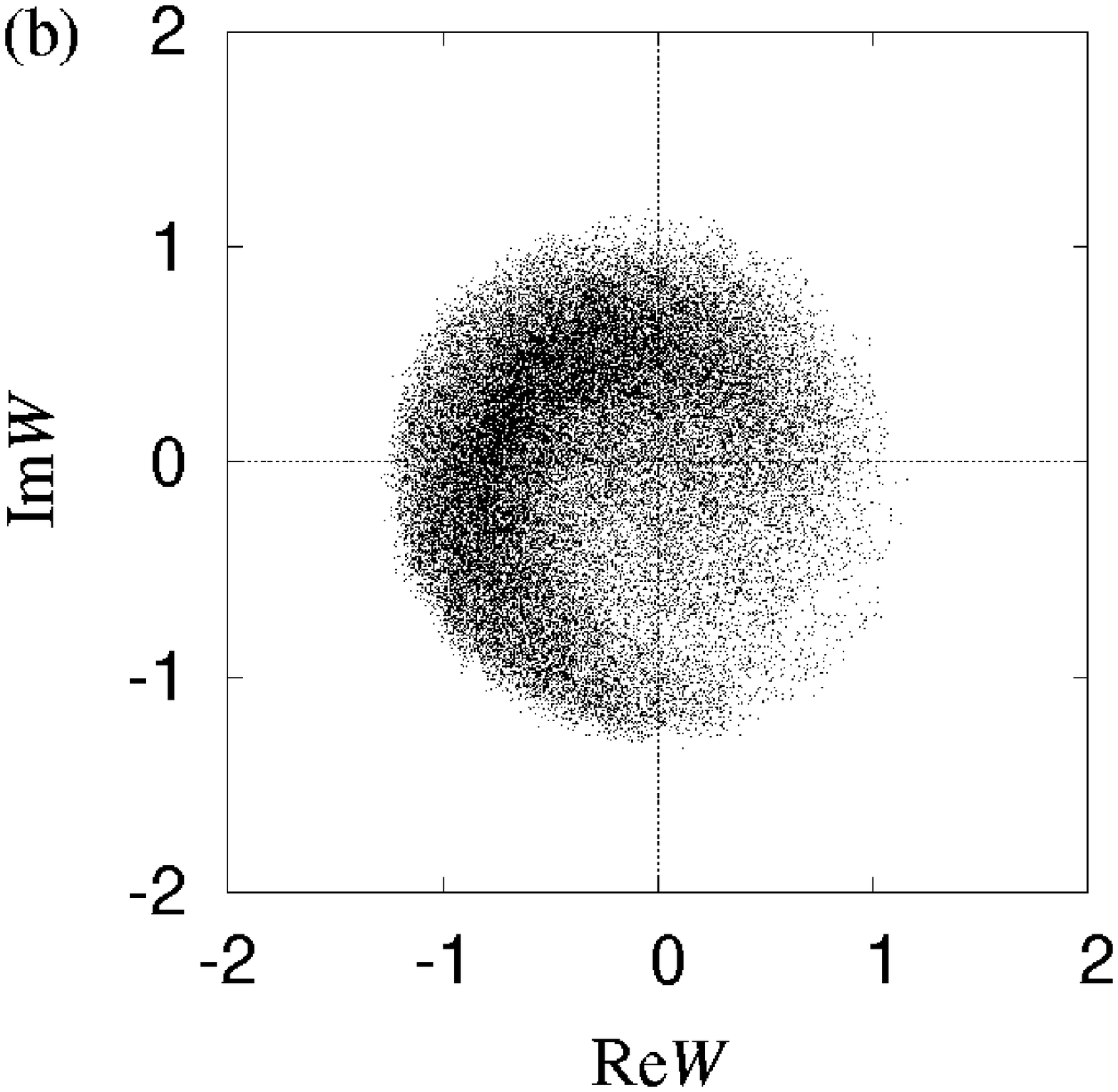}
\caption{Phase portraits for $\mu=0.1$ (a) and $\mu=0.3$ (b)
at a given time. The system size is $N=40000$.}
\label{fig:phase}
\end{figure}

\begin{figure}
\centering
\includegraphics[width=8cm]{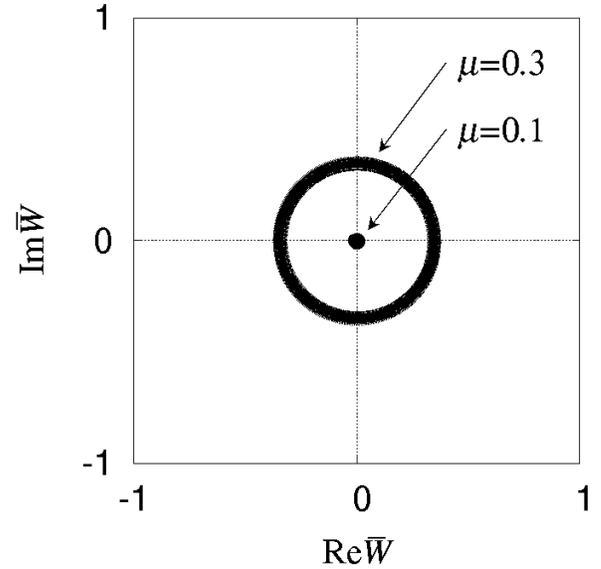}
\caption{Trajectories of the mean field in the complex plane
for $\mu=0.1$ and $\mu=0.3$. The mean field is fluctuating
around the origin when $\mu=0.1$, while it forms a closed
orbit with small amplitude fluctuations when $\mu=0.3$.
The fluctuations come from the finiteness of the system size.
The system size is $N=40000$.}
\label{fig:orbit}
\end{figure}

In Fig.~\ref{fig:stp}, two space-time plots of the
complex amplitude modulus for $\mu=0.1$ (weaker feedback)
and $\mu=0.3$ (stronger feedback) are indistinguishable.
In Fig.~\ref{fig:phase}, however, two phase portraits
in the complex amplitude plane each obtained for $\mu=0.1$
and $\mu=0.3$ are contrasted with each other.
While the distribution of the representative points of the local
oscillators looks almost isotropic for the case of $\mu=0.1$,
implying the absence of collective oscillations, such symmetry is
obviously lost when $\mu=0.3$, implying the existence of collective
oscillations.
Qualitative difference between the two situations is further confirmed
from Fig.~\ref{fig:orbit} where a trajectory of the mean field
over a long time at each value of $\mu$ is displayed.
It is clear that when the feedback is weak the mean field is non-oscillatory,
simply fluctuating (presumably due to the finite-size effects described above)
around the origin, whereas for stronger feedback the same quantity clearly
exhibits a closed orbit with some amplitude fluctuation again due to the
finite-size effects.
Thus, if there is a transition somewhere between these $\mu$-values, it is
presumably characterized by a noisy Hopf bifurcation.

The nonlinear Langevin equation~(\ref{eq:le}) actually predicts
the occurrence of a noisy Hopf bifurcation.
From the symmetry of our system and the smallness in amplitude of the collective
oscillation near its onset, the nonlinear effects ${\cal N}(\bar{W})$ in
Eq.~(\ref{eq:le}) will be dominated by a cubic term.
Then the mean field obeys a noisy Stuart-Landau equation
\begin{equation}
\dot{\bar{W}}
=\left[\left(\mu-\mu_{\rm c}\right)+i\omega\right]\bar{W}
-\left(\eta+i\alpha\right)\left|\bar{W}\right|^2\bar{W}+f(t),
\label{eq:sl}
\end{equation}
where the coefficients $\mu_{\rm c}$, $\omega$, $\eta(>0)$, and $\alpha$ are
real and depend generally on $c_1$ and $c_2$.
Since the above equation can be handled analytically, it would be interesting
to compare some of the results from its analysis with our direct numerical
simulations on Eq.~(\ref{eq:cgl}) and its two-dimensional extension,
which are the subjects of Secs.~\ref{sec:results} and \ref{sec:two},
respectively.

\section{Predicted critical behavior and comparison with numerical results}
\label{sec:results}

The nonlinear Langevin equation~(\ref{eq:sl}) is equivalent
with the Fokker-Planck equation of the following form
~\cite{ref:risken89,ref:kampen81}:
\begin{align}
\frac{\partial}{\partial t} P\left(r,\phi,t\right)
&=\frac{1}{r}\frac{\partial}{\partial r}
\left\{\left[-\left(\mu-\mu_{\rm c}\right)r^2+\eta r^4\right]P
+\Gamma r \frac{\partial}{\partial r}P\right\} \nonumber \\
&+\left[\left(-\omega+\alpha r^2\right)
\frac{\partial}{\partial \phi}P
+\frac{\Gamma}{r^2}\frac{\partial^2}{\partial \phi^2}P\right].
\label{eq:fp}
\end{align}
Here $P(r,\phi,t)$ is the probability density for the amplitude $r$
and the phase $\phi$ of the mean field at time $t$.
The above equation admits a stationary solution independent of $\phi$
given by
\begin{equation}
P_{\rm st}\left(r\right) = \exp \left[-\frac{\eta}{4\Gamma}
\left(r^2-\frac{\mu-\mu_{\rm c}}{\eta}\right)^2\right].
\label{eq:st1}
\end{equation}
Various moments of $r$ defined by
\begin{equation}
\left\langle r^n\right\rangle =
\frac{\displaystyle \int_0^\infty r^n P_{\rm st}\left(r\right)r dr}
{\displaystyle \int_0^\infty P_{\rm st}\left(r\right)r dr},
\label{eq:rn}
\end{equation}
and those of the fluctuation $\delta r\equiv r-\langle r\rangle$
can be calculated~\cite{ref:risken65}.
In particular, in the limit of weak random force, the mean field amplitude
$\langle r\rangle$ near the transition is found to depend on $\mu-\mu_{\rm c}$ as
\begin{equation}
\left\langle r\right\rangle=
\begin{cases}
A\left(\mu-\mu_{\rm c}\right)^{1/2} &
\quad\left(\mu>\mu_{\rm c}\right), \\
0 & \quad\left(\mu<\mu_{\rm c}\right),
\end{cases}
\label{eq:r1}
\end{equation}
where $A$ is a constant.
Similarly, the mean field fluctuation $\langle (\delta r)^2\rangle$ is given by
\begin{equation}
\left\langle(\delta r)^2\right\rangle\propto\left|\mu-\mu_{\rm c}\right|^{-1},
\label{eq:dr2}
\end{equation}
which holds for $\mu\gtrless\mu_{\rm c}$.
It is clear that the critical exponents associated with $\langle r\rangle$
and $\langle(\delta r)^2\rangle$ obey the classical law or the mean field theory,
reflecting the fact that the transition is caused by the applied mean field and
not by the development of local order to a macroscopic scale.
Still the transition is in some sense statistical in nature unlike bifurcations
in deterministic dynamical systems.
This is because the loss of long-range order is solely due to turbulent fluctuations,
so that a full theoretical understanding of the transition phenomenon would be
impossible without statistical mechanics of chemical turbulence.

\begin{figure}
\centering
\includegraphics[width=8cm]{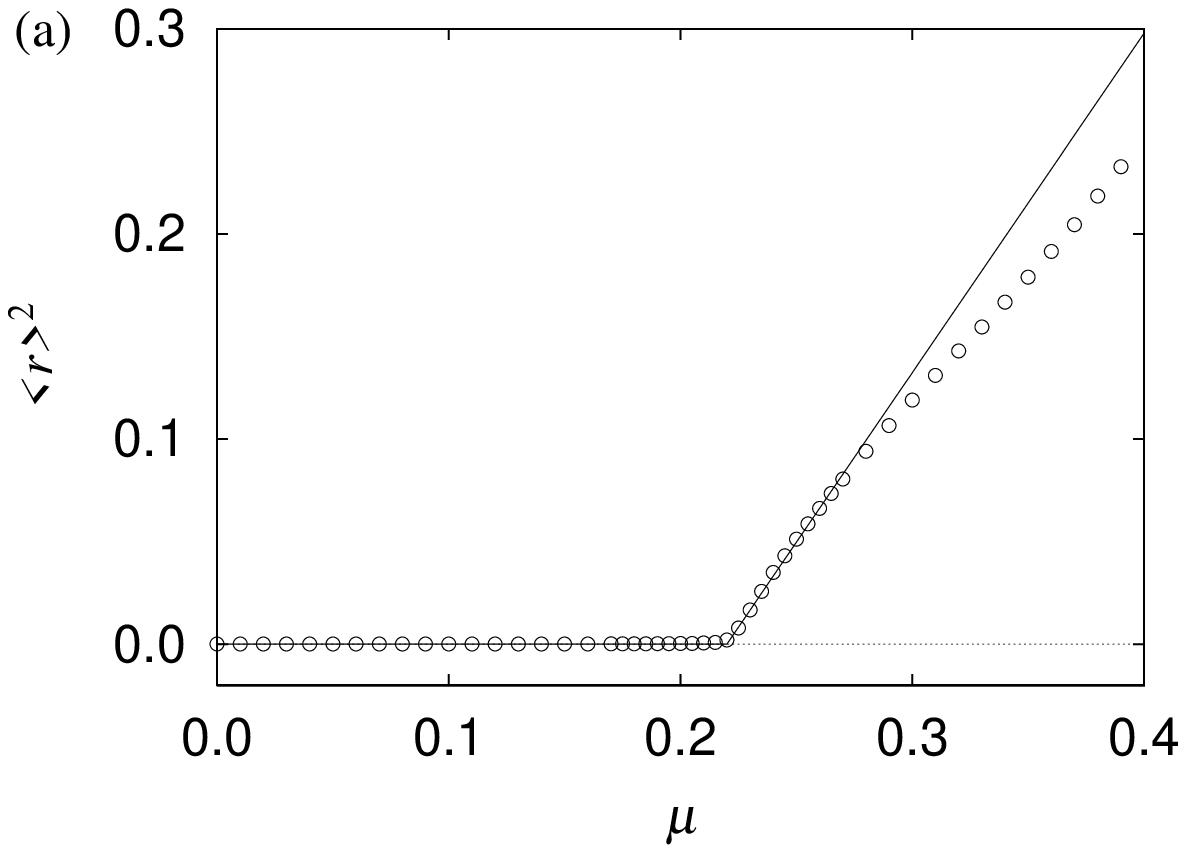} \\
\includegraphics[width=8cm]{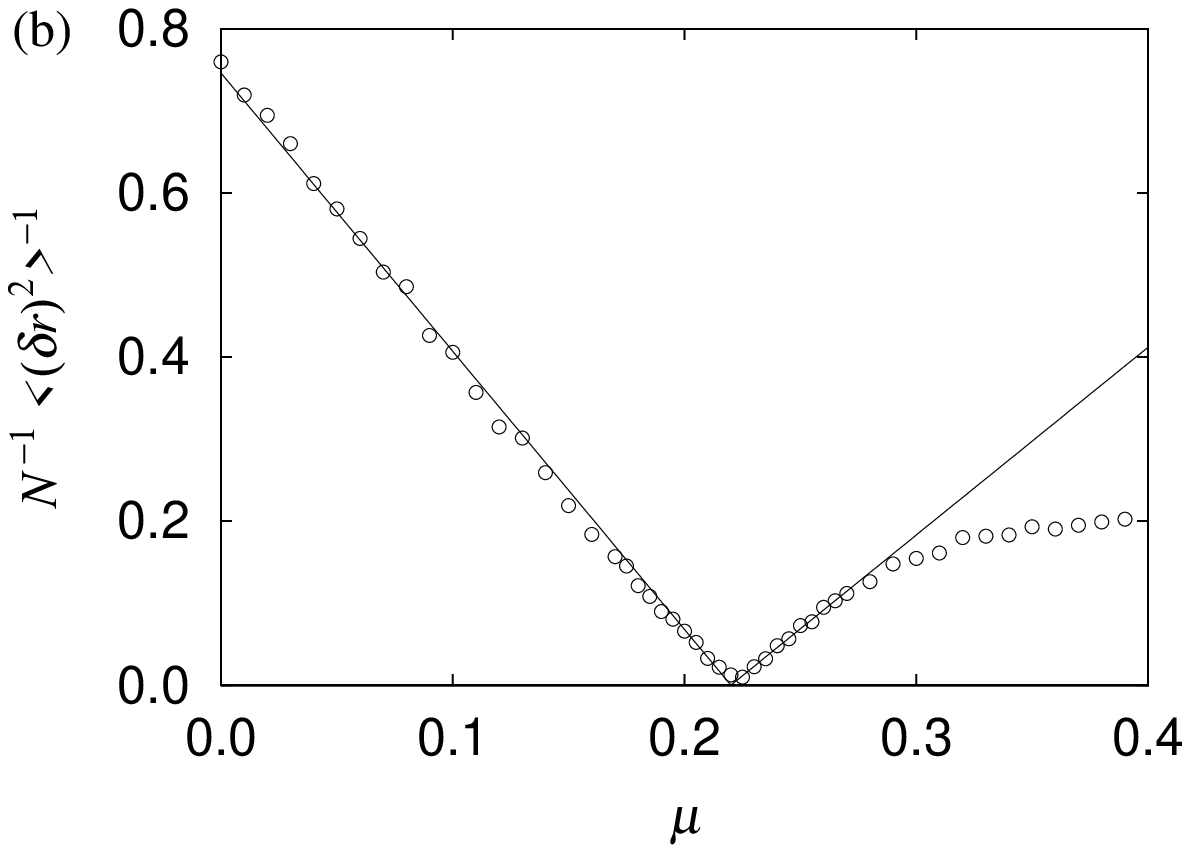}
\caption{$\langle r\rangle^2$ vs.\ $\mu$ (a).
$N^{-1}\langle(\delta r)^2\rangle^{-1}$ vs.\ $\mu$ (b).
The open circles and the solid lines are the numerical data
and the theoretical curves, respectively. In each of (a) and (b),
a suitable scale factor is used as an adjustable parameter to
achieve a good fitting between the theory and numerical experiments.
The system size is $N=200000$.}
\label{fig:exponent}
\end{figure}

Long-time averages of $r$ and $(\delta r)^2$ were obtained as a function
of the feedback intensity from numerical simulation of Eqs.~(\ref{eq:cgl})
and (\ref{eq:gf}) with $N=200000$, and the results are shown in
Figs.~\ref{fig:exponent}(a) and (b), respectively.
By assuming that the long-time averages are identical with ensemble averages,
these numerical data were fitted with the theoretical curves given by
Eqs.~(\ref{eq:r1}) and (\ref{eq:dr2}), respectively, where a scale
factor is used as the only adjustable parameter.
Note that $\mu_{\rm c}$ is not an adjustable  parameter, but is a constant
given by Eq.~(\ref{eq:muc}).

\begin{figure}
\centering
\includegraphics[width=8cm]{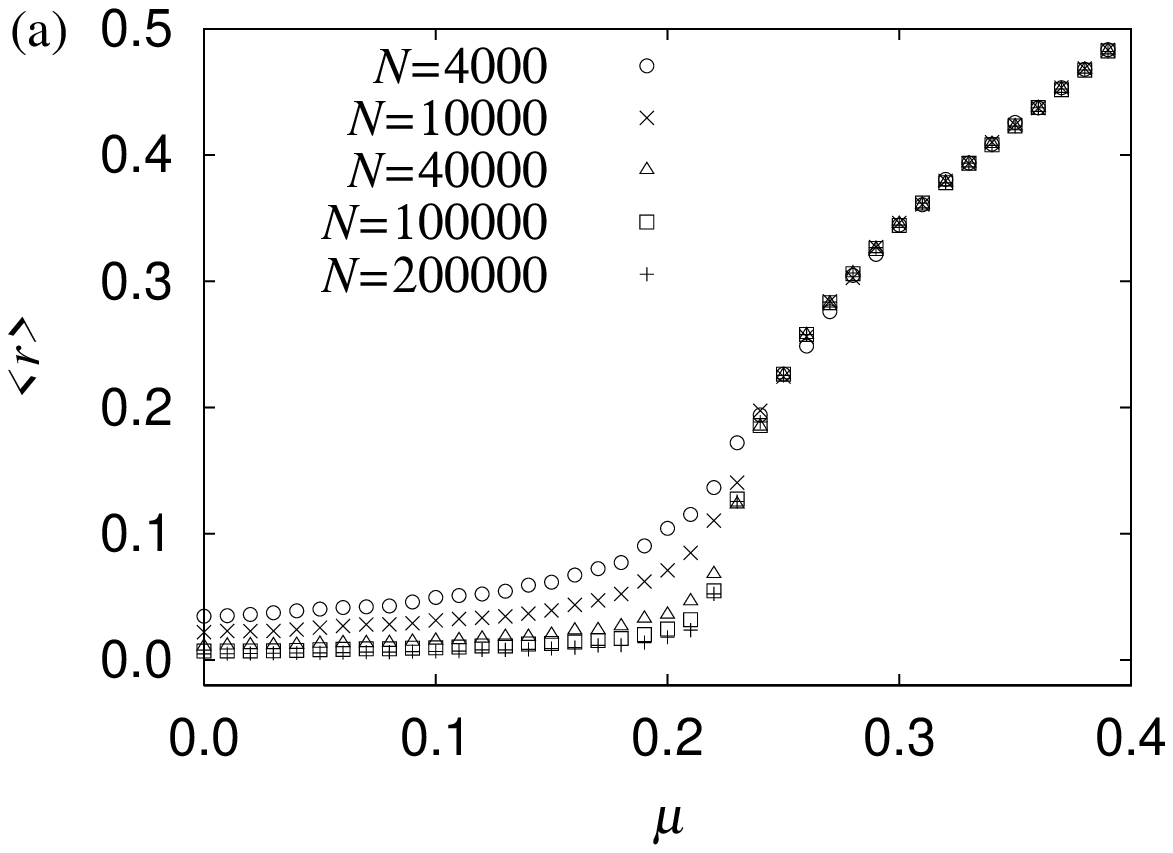} \\
\includegraphics[width=8cm]{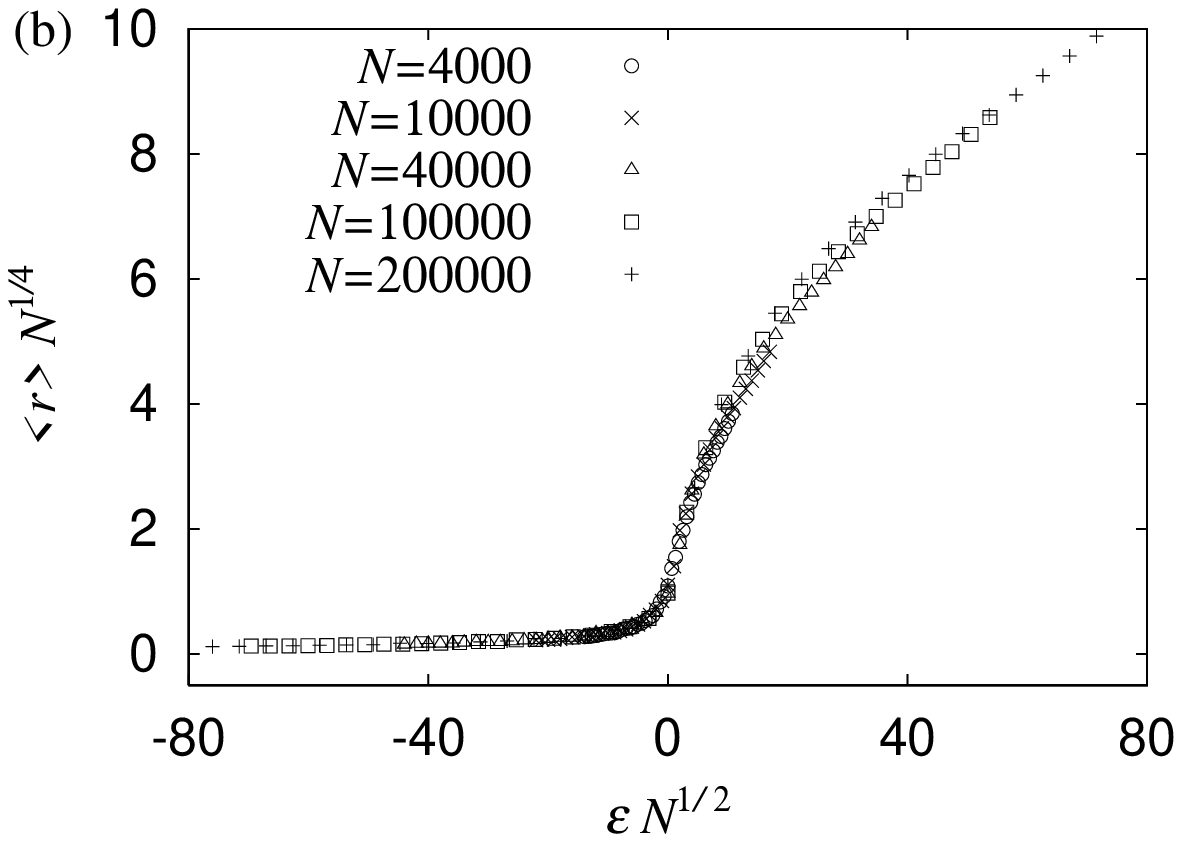}
\caption{$\langle r\rangle$ vs.\ $\mu$ for some different values of $N$ (a).
Rescaling of the data in (a) according to the finite-size scaling
given by Eq.~(\ref{eq:fss}) produces a universal curve independent
of $N$ (b).}
\label{fig:finite}
\end{figure}

In obtaining Eq.~(\ref{eq:r1}) for the mean field amplitude,
we considered the limit of weak random force.
Let our argument be generalized to include the dependence of
$\langle r\rangle$ on the noise intensity as well as on $\mu-\mu_{\rm c}$.
Because the origin of the random force driving the mean field is the
finiteness of the system size, the noise intensity should be inversely
proportional to the system size, or
\begin{equation}
\Gamma=\frac{D}{N},
\label{eq:fse}
\end{equation}
where $D$ is a constant independent of $N$.
One may thus write the stationary distribution in the form
\begin{equation}
P_{\rm st}(r)=\exp\left[-\frac{N\eta}{4D}
\left(r^2-\frac{\varepsilon}{\eta}\right)^2\right],
\label{eq:st2}
\end{equation}
where $\varepsilon=\mu-\mu_{\rm c}$. Applying the finite-size scaling
law developed in Ref.~\cite{ref:pikovsky99} to the average amplitude
of the mean field, we obtain a scaling form
\begin{equation}
\left\langle r\right\rangle=N^{-1/4} F\left(\varepsilon N^{1/2}\right),
\label{eq:fss}
\end{equation}
where $F$ is a function (called the scaling function) depending on
$\varepsilon$ and $N$ only through $\varepsilon N^{1/2}$.
Equation~(\ref{eq:fss}) is a generalization of Eq.~(\ref{eq:r1}).
Numerically calculated $\langle r\rangle$ for various $\varepsilon$ and $N$
confirms this scaling law.
Figure~\ref{fig:finite}(a) shows the dependence of the long-time
average of the mean field amplitude on feedback intensity for some
different values of $N$.
As is seen from Fig.~\ref{fig:finite}(b), all these data come to lie
on an identical universal curve after the rescalings of $\langle r\rangle$
and $\varepsilon$ by $N^{1/4}$ and $N^{1/2}$, respectively.
In this way, the finite-size scaling law~(\ref{eq:fss}) is confirmed,
providing unmistakable evidence for a phase transition.

\section{Two-dimensional case} \label{sec:two}

\begin{figure}
\centering
\includegraphics[width=8cm]{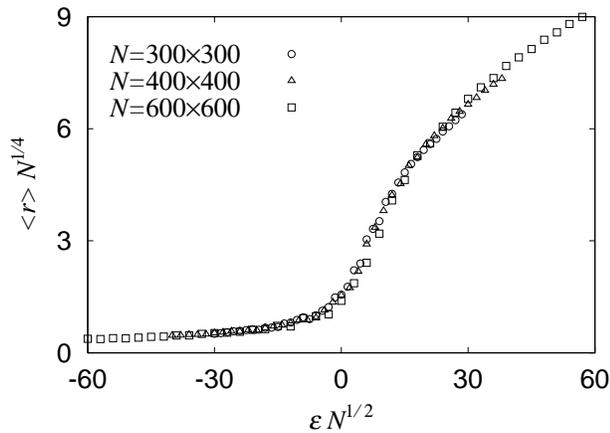}
\caption{Similar to Fig.~\ref{fig:finite}(b) but in two space dimensions.
Parameter values are the same as in the one-dimensional case, i.e., $c_1=2.0$
and $c_2=-2.0$, which gives $\mu_{\rm c}\simeq 0.10$.
This value is used in rescaling of the numerical data.}
\label{fig:two}
\end{figure}

The one-dimensional reaction-diffusion systems which we have
numerically studied in Secs.~\ref{sec:langevin} and
\ref{sec:results} are not very realistic.
Especially, surface chemical reactions, such as catalytic
CO oxidation on Pt, occur in two dimensions.
As mentioned in the foregoing sections, the transition considered here
is the mean field type, so that the nature of the transition is expected
to be the same as that in one-dimensional systems.
We carried out numerical simulations on the two-dimensional complex
Ginzburg-Landau equation with global feedback, and obtained a clear
indication of transition, although the corresponding value of
$\mu_{\rm c}$ is considerably smaller than  that of the one-dimensional
case under the same parameter condition.
Figure~\ref{fig:two} summarizes numerical results for
various $\varepsilon$ and $N$, exhibited in a similar manner
to Fig.~\ref{fig:finite}(b), i.e., in the form of
$\langle r\rangle N^{1/4}$ vs.\ $\varepsilon N^{1/2}$
for different values of $N$.
These data form almost an identical curve again, which we take as
evidence for a transition similar to that in one-dimensional systems.

\section{Concluding remarks} \label{sec:remarks}

Preceding the transition at which the turbulence is completely
suppressed and uniform oscillations set in, a different type of
transition characterized by the emergence of collective oscillations
was shown to exist in the one- and two- dimensional complex
Ginzburg-Landau equations with global feedback.
The transition is well described phenomenologically with the noisy
Stuart-Landau equation governing the mean field.
Since the noise there comes from the finite-size effects, the transition
becomes infinitely sharp in the limit of infinite system size.
The critical exponents of this transition obey the mean field theory
because the origin of cooperativity is nothing but the mean field produced
by the global feedback.
This also gives the reason why spatial dimension one is sufficient for
giving rise to the transition.
We also confirmed that the two-dimensional complex Ginzburg-Landau equation
with global feedback which is more realistic also exhibits a transition of
the same type.

Throughout the present paper, our analysis was confined to the case
that the feedback intensity $\mu$ is real.
From our ongoing study, it is being confirmed that a transition of the same
nature persists over some range of complex $\mu$.
The nonlinear Langevin equation~(\ref{eq:le}) also seems to remain valid.
Unlike the case of real $\mu$, however, the effective damping coefficient
$\gamma$ appearing in the Langevin equation~(\ref{eq:le}) seems to
depend on $\mu$, which raises an interesting theoretical problem to be
tackled in the future.

Our analysis~\cite{ref:kawamura06} also suggests that the realistic dynamical
model for the catalytic CO oxidation~\cite{ref:kim01,ref:bertram03-2}
under experimentally accessible parameter conditions also exhibit a
similar transition.
We strongly hope for its experimental verification.

\begin{acknowledgments}
The authors thank H.~Nakao for useful discussions.
Numerical computation of this work was carried out with the Computer
Facility at Yukawa Institute for Theoretical Physics, Kyoto University.
\end{acknowledgments}


\end{document}